\documentclass[twocolumn,showpacs,preprintnumbers,amsmath,amssymb]{revtex4-1}
\usepackage{graphicx}
\begin{document}

\renewcommand{\theequation}{\thesection.\arabic{equation}}

\newcommand{\re}{\mathop{\mathrm{Re}}}

\newcommand{\be}{\begin{equation}}
\newcommand{\ee}{\end{equation}}
\newcommand{\bea}{\begin{eqnarray}}
\newcommand{\eea}{\end{eqnarray}}

\title{Spacetime averaging of exotic singularity universes.}

\author{Mariusz P. D\c{a}browski}
\email{mpdabfz@wmf.univ.szczecin.pl}
\affiliation{\it Institute of Physics, University of Szczecin, Wielkopolska 15, 70-451 Szczecin, Poland}
\affiliation{\it Copernicus Center for Interdisciplinary Studies,
S{\l }awkowska 17, 31-016 Krak\'ow, Poland}

\date{\today}

\input epsf

\begin{abstract}
Taking a spacetime average as a measure of the strength of singularities we show that big-rips (type I) are stronger than big-bangs. The former have infinite spacetime averages while the latter have them equal to zero. The sudden future singularities (type II) and $w-$singularities (type V) have finite spacetime averages. The finite scale factor (type III) singularities for some values of the parameters may have an infinite average and in that sense they may be considered stronger than big-bangs.

\end{abstract}


\maketitle

\section{Introduction}
\setcounter{equation}{0}

Following the observational evidence for the universe acceleration \cite{supernovae} some new types of cosmological singularities were proposed
(cf. Refs. \cite{nojiri,APS2010}). These were
first of all big-rips, due to a phantom matter \cite{phantom}, and then sudden future singularities \cite{SFS}, finite scale factor (type III)
singularities \cite{nojiri}, big separation (type IV) \cite{nojiri}, and $w-$singularities (type V) \cite{wsing}.

Phantom dark energy leads to a big-rip singularity in which all the matter is dissociated by gravity in a large and a dense universe.
This behavior is of course different from the standard picture of cosmic evolution which allows big-bang or big-crunch types of
singularities only. Standard dark energy models are based on the matter which violates the strong energy condition ($\varrho + p \geq 0$
and $\varrho + 3p \geq 0$). Phantom matter, on the other hand,  violates all the remaining energy conditions too, i.e., the null
($\varrho + p \geq 0$), weak ($\varrho \geq 0$ and $\varrho + p \geq 0$),  and dominant energy
($\varrho \geq 0$, $-\varrho  \leq p \leq \varrho $) (here $c=1$, $\varrho$ is the energy density, and $p$ is the pressure).
A sudden future singularity model violates only the dominant energy condition, its generalized version known as a generalized sudden
future singularity model \cite{SFS}, does not violate any of the energy conditions and this is also true for a big-separation and a $w$-singularity.
The exotic singularities are characterized by a blow-up of all or some of the appropriate physical quantities such as: the scale factor,
the energy density, the pressure, and the barotropic index (for a review see Ref. \cite{APS2010}). It is interesting that these singularities
may be inspected observationally by using the higher-order characteristics of the expansion of the universe \cite{plb05} known as
statefinders \cite{statef}. In particular, sudden future singularities (which include the so-called big-brakes \cite{big-brake})
have been tested against supernovae data \cite{obsSFS}.

One of the problems in cosmology is that for exotic singularity models one cannot tight the standard tools known from relativity
such as the energy conditions with the presence of singularities. A suggestion to formulate higher-order energy conditions
to deal with the problem has been made in Ref. \cite{plb05}. The task of this paper is to exercise yet another tool which is the
spacetime averaging of the singular cosmological quantities.

\section{Spacetime averaging}
\setcounter{equation}{0}

According to Raychaudhuri (Ref. \cite{raych98}) one is always able to take an average of any physical or kinematic scalar quantity $\chi$ over the entire (open)
spacetime in the form
\be
<\chi> \equiv \left[ \frac{\int_{-x_0}^{x_0} \ldots \int_{-x_3}^{x_3} \chi \sqrt{-g} d^4 x}{\int_{-x_0}^{x_0} \ldots \int_{-x_3}^{x_3}
\sqrt{-g} d^4 x}\right]_{\lim_{x_0, \ldots x_3 \to \infty}}.
\label{averchi}
\ee
Using (\ref{averchi}), Raychaudhuri made a claim that any singularity-free non-rotating universe which is open in all
directions had the spacetime average of the stress-energy tensor invariants, including the energy density, equal to zero. By open in all
directions he meant that the ratio of the 3-volume hypersurfaces of any type (spacelike, timelike) to the 4-volume of spacetime vanishes, i.e.,
\be
\frac{\int\int\int \sqrt{\mid ^3g \mid} dx^idx^jdx^k}{\int\int\int\int \sqrt{-g}d^4x} = 0~~,
\ee
where $i, j, k$ are different and can be both spatial and temporal coordinates.
That way he wanted to paradigm a non-singular model of Senovilla \cite{senov} which apparently had the average (\ref{averchi}) vanishing.
However, it emerged that the problem is subtler and that to relate the vanishing of spacetime average to a lack of spacetime singularity
is not very conclusive. The situation is analogous to what we face when we try to tight the energy conditions (which in fact follow from
famous Raychaudhuri equation anyway) with the appearance of singularities. For example, even if all the energy conditions
are fulfilled, the big-bang singularity is possible in standard cosmology. On the other hand, for phantom, all the energy conditions are
violated and the big-rip singularity appears \cite{plb05}. Let us remind that both big-bang and big-rip singularities are geodesically incomplete \cite{leonardo1} so that they are the true singularities according to singularity theorems of Hawking and Penrose. As it was already mentioned, for exotic singularities \cite{APS2010}, which are
generally weak singularities (geodesically complete) \cite{leonardo1}, it might be that these singularities lead to violation of some of the energy conditions only.

Similar way to average, though applied to spatially inhomogeneous universes, was proposed by Buchert \cite{buchert} who defined
a spatial average of a quantity ${\cal A}$ as
\be
< {\cal A} > \equiv \frac{\int\int\int {\cal A} \sqrt{\mid ^3g \mid} d^3x}{\int\int\int \sqrt{\mid ^3g \mid} d^3x}~~.
\ee
The method was applied, among others, for the Tolman universes \cite{tolman}. However, since in our case only the homogeneous models
are studied, we will use the spacetime averaging (\ref{averchi}) further.

Yet another approach to differentiate between singular and non-singular cosmologies was applied in Ref. \cite{canmeasure}, where the method of canonical (Liouville-Henneaux-Gibbons-Hawking-Stewart) measure was applied. It was shown in Ref. \cite{page} that this measure was finite for non-singular Friedmann cosmologies with a minimally coupled scalar field and a positive cosmological constant. This also happens to be finite for the models with a minimally coupled scalar field and a positive $\Lambda-$term allowing a big-bang and a big-crunch. On the other hand, it was also shown in Ref. \cite{page} that the models which expand forever had an infinite canonical measure.

Coming back to the Raychaudhuri claim made in Ref. \cite{raych98}, it was easy to show \cite{saa} that his claim was not true even for a flat Friedmann universe.
In fact, for such a model ($\sqrt{-g} = a^3(t)$, where $a(t)$ is the scale factor) an average acceleration of the universe vanishes for dust and radiation models. Thus, it is not correct to tight the appearance of singularities to a vanishing of the spacetime average of the physical and kinematical scalars.

In this paper we will discuss an issue of spacetime averaging of the standard and exotic singularity Friedmann models to support this claim in the context of a more general set of models which violate only some or none of the energy conditions.

\section{Application to standard and exotic singularity models}
\setcounter{equation}{0}

Let us first take an average acceleration scalar $\chi = \theta_{,\mu}u^{\mu}$ according to (\ref{averchi}) for a flat Friedmann model which reads as
\be
<\dot{\theta}> = \lim_{\genfrac{}{}{0pt}{}{t_0 \to 0}{t_1 \to \infty}} \frac{3\int_{t_0}^{t_1}a^3\left(\frac{\ddot{a}}{a} - \frac{\dot{a}^2}{a^2} \right)dt}{\int_{t_0}^{t_1}a^3 dt}.
\label{thetaav}
\ee
Assuming the standard barotropic equation of state $p=(\gamma - 1)\varrho = w\varrho$ ($\gamma$ is the barotropic index, more recently written down as $w$), we basically have three cases: $\gamma > 0$ (standard matter), $\gamma = 0$ (de Sitter), $\gamma < 0$ (phantom). For $\gamma = 0$ we have $a(t) = a_0 e^{H_0 t}$ which gives $<\dot{\theta}> = 0$ no matter what are the time limits. For $\gamma > 0$ we have $a(t) = t^{2/3\gamma}$ and so for $\gamma \neq 2$
\bea
<\dot{\theta}>_{stand} &=& \lim_{\genfrac{}{}{0pt}{}{t_0 \to 0}{t_1 \to \infty}} - \frac{2}{\gamma} \frac{\int_{t_0}^{t_1} t^{2\left(\frac{1}{\gamma} - 1 \right)} dt}{\int_{t_0}^{t_1} t^{\frac{2}{\gamma}}dt} \\
&=& \lim_{\genfrac{}{}{0pt}{}{t_0 \to 0}{t_1 \to \infty}} - \frac{2}{\gamma} \frac{\frac{2}{\gamma} + 1}{\frac{2}{\gamma} - 1} \frac{{t_1}^{\frac{2}{\gamma}-1} - {t_0}^{\frac{2}{\gamma}-1}}{{t_1}^{\frac{2}{\gamma}+1} - {t_0}^{\frac{2}{\gamma}+1}} \to 0 \, \nonumber
\label{thetaavstan}
\eea
and for $\gamma = 2$ (stiff-fluid)
\be
<\dot{\theta}>_{stiff} = \lim_{\genfrac{}{}{0pt}{}{t_0 \to 0}{t_1 \to \infty}} 2 \frac{\ln{t_1} - \ln{t_0}}{t_1^2 - t_0^2}.
\label{thetaavstiff}
\ee
Apparently, the limit (\ref{thetaavstiff}) is singular, but after calculating it carefully, one sees that it also gives zero so that the limit of (\ref{thetaav}) is zero for an arbitrary, but positive value of the barotropic index $\gamma >0$. However, the situation is entirely different for phantom $\gamma < 0$. Defining $\gamma = -\mid \gamma \mid < 0$ for a phantom case, one has that the scale factor $a(t) = t^{-2/3 \mid \gamma \mid}$, and so the integral (\ref{thetaav}) for $\mid \gamma \mid \neq 2$ reads as
\bea
<\dot{\theta}>_{ph} &=& \lim_{\genfrac{}{}{0pt}{}{t_0 \to 0}{t_1 \to \infty}} \frac{2}{\mid \gamma \mid} \frac{\int_{t_0}^{t^1} t^{-2\left(1+ \frac{1}{\gamma} \right)} dt}{\int_{t_0}^{t_1} t^{-\frac{2}{\mid \gamma \mid}}dt}
\label{thetaavphan} \\
&=& \lim_{\genfrac{}{}{0pt}{}{t_0 \to 0}{t_1 \to \infty}} \frac{2}{\mid \gamma \mid} \frac{\frac{2}{\mid \gamma \mid} - 1}{\frac{2}{\mid \gamma \mid} + 1} \frac{{t_1}^{-\frac{2}{\mid \gamma \mid}-1} - {t_0}^{-\frac{2}{\mid \gamma \mid}-1}}{{t_1}^{-\frac{2}{\mid \gamma \mid}+1} - {t_0}^{-\frac{2}{\mid \gamma \mid}+1}} \to \infty \, \nonumber
\eea
and for $\mid \gamma \mid = 2$ it is
\be
<\dot{\theta}>_{ph,st} = \lim_{\genfrac{}{}{0pt}{}{t_0 \to 0}{t_1 \to \infty}} -\frac{1}{2} \frac{\frac{1}{t_1^2} - \frac{1}{t_0^2}}{\ln{t_1} - \ln{t_0}}.
\label{thetaavphst}
\ee
One should mention that the limits $t_0$ and $t_1$ in (\ref{thetaavstan})-(\ref{thetaavstiff}) could have been taken for a second branch solution for negative times and so we would have $t_0 \to - \infty$,
$t_1 \to 0$. However, this dual solution also gives the limit for $<\dot{\theta}>$ to be zero, and it does not change the whole picture. The same is true for phantom. Usually, a big-rip singularity is considered to take place in the end of the evolution $(t_1 \to 0, t_0 \to \infty)$, but we chose in (\ref{thetaavphan})-(\ref{thetaavphst}) that the evolution starts at a big-rip. In conclusion, we can see a large difference between the behaviour of the acceleration scalar in a standard and a phantom case. For the former, an average acceleration vanishes, while for the latter, it diverges. The conclusion is interesting, since it may suggest that the phantom cosmological models possess stronger singularities (big-rips) than standard big-bang models.

Similar conclusion follows from the averaging of the energy density $\varrho$ and the pressure $p$ ($8\pi G = 1$)
\be
<p> = - \lim_{\genfrac{}{}{0pt}{}{t_0 \to 0}{t_1 \to \infty}} \frac{\int_{t_0}^{t_1}a^3\left(2\frac{\ddot{a}}{a} + \frac{\dot{a}^2}{a^2} \right)dt}{\int_{t_0}^{t_1}a^3 dt}~~,
\label{peav}
\ee
and
\be
<\varrho> = \lim_{\genfrac{}{}{0pt}{}{t_0 \to 0}{t_1 \to \infty}} \frac{3\int_{t_0}^{t_1}a^3\left(\frac{\dot{a}^2}{a^2}\right)dt}
{\int_{t_0}^{t_1}a^3 dt}~~.
\label{roav}
\ee
For a barotropic perfect fluid we obtain the same integrals as for average acceleration up to constants, i.e.,
\bea
<p>_{stand} &=& \lim_{\genfrac{}{}{0pt}{}{t_0 \to 0}{t_1 \to \infty}} - \frac{4}{\gamma} \left(\frac{1}{\gamma} - 1 \right)\frac{\int_{t_0}^{t_1} t^{2\left(\frac{1}{\gamma} - 1 \right)} dt}{\int_{t_0}^{t_1} t^{\frac{2}{\gamma}}dt} ,\nonumber
\label{peav1}\\
<\varrho>_{stand} &=& \lim_{\genfrac{}{}{0pt}{}{t_0 \to 0}{t_1 \to \infty}} - \frac{4}{3\gamma^2} \frac{\int_{t_0}^{t_1} t^{2\left(\frac{1}{\gamma} - 1 \right)} dt}{\int_{t_0}^{t_1} t^{\frac{2}{\gamma}}dt} \, \nonumber
\label{roav1}
\eea
for standard matter, and
\bea
<p>_{ph} &=& \lim_{\genfrac{}{}{0pt}{}{t_0 \to 0}{t_1 \to \infty}} - \frac{4}{\mid \gamma \mid} \left(\frac{1}{\mid \gamma \mid} + 1 \right)\frac{\int_{t_0}^{t_1} t^{-2\left(\frac{1}{\mid \gamma \mid} + 1 \right)} dt}{\int_{t_0}^{t_1} t^{-\frac{2}{\mid \gamma \mid}}dt} ,\nonumber
\label{peav2}\\
<\varrho>_{ph} &=& \lim_{\genfrac{}{}{0pt}{}{t_0 \to 0}{t_1 \to \infty}} - \frac{4}{3\gamma^2} \frac{\int_{t_0}^{t_1} t^{-2\left(\frac{1}{\mid \gamma \mid} + 1 \right)} dt}{\int_{t_0}^{t_1} t^{-\frac{2}{\mid \gamma \mid}}dt} \, \nonumber
\label{roav2}
\eea
for phantom.

Now, we consider a pressure singularity model which starts with a big-bang, then continues to a sudden future singularity of pressure, and due to its geodesic completeness \cite{leonardo1} continues to a big-crunch as follows \cite{adam}:
\be
a_L (t) = a_s \left[\delta + \left(1 + \frac{t}{t_B} \right)^m \left(1 - \delta \right) - \delta \left(- \frac{t}{t_B} \right)^n \right]
\label{aL}
\ee
with $t_B < 0$ - a big-bang time, $a_L(-t_B)=0$, $t=0$ a sudden future singularity time, $a_{L}(0)=a_{R}(0) = a_s$:
\be
a_R (t) = a_s \left[\delta + \left(1 - \frac{t}{t_C} \right)^m \left(1 - \delta \right) - \delta \left(\frac{t}{t_C} \right)^n \right]
\label{aR}
\ee
with $t_C > 0$ - a big-crunch time, $a_R(t_C)=0$, and $a_s, \delta, m=$ const., $1<n<2$. Near to a pressure singularity $t \to 0$ (\ref{aL}) and (\ref{aR}) are approximated by
\bea
a_L & \approx & a_s \left[ 1 + \frac{m}{t_B} \left(1 - \delta \right) t \right], \\
a_R & \approx & a_s \left[ 1 - \frac{m}{t_C} \left(1 - \delta \right) t \right].
\eea
The first and the second derivatives of the scale factors (\ref{aL})-(\ref{aR}) on the left and right of sudden singularity are given by
\be
\dot{a}_L (t) = a_s \left[\frac{m}{t_B} \left(1 + \frac{t}{t_B} \right)^{m-1} \left(1 - \delta \right) + \delta \frac{n}{t^n_B} \left(- t \right)^{n-1} \right]~~,
\label{aLd}
\ee
\be
\dot{a}_R (t) = a_s \left[-\frac{m}{t_C} \left(1 - \frac{t}{t_C} \right)^{m-1} \left(1 - \delta \right) + \delta \frac{n}{t^n_C} \left(t \right)^{n-1} \right]~~,
\label{aRd}
\ee
\be
\frac{\ddot{a}_L}{a_s} = \frac{m(m-1)(1-\delta)}{t^2_B} \left(1 + \frac{t}{t_B} \right)^{m-2} - \frac{\delta n(n-1)}{t^n_B} \left(- t \right)^{n-2}~~,
\label{aLdd}
\ee
\be
\frac{\ddot{a}_R}{a_s} = \frac{m(1-m)(1-\delta)}{t^2_C} \left(1 - \frac{t}{t_C} \right)^{m-2} + \frac{\delta n(n-1)}{t^n_C} t^{n-2}~~.
\label{aRdd}
\ee
Only the last terms in (\ref{aLdd})-(\ref{aRdd}) blow up for $1<n<2$ at $t=0$, so that to calculate $<\dot{\theta}>$, $<\dot{p}>$ and $<\dot{\varrho}>$ which reflect the effect of a sudden singularity only (we have already shown that average over the big-bang for standard matter is zero), one may also use the last terms of (\ref{aL})-(\ref{aR}) and (\ref{aLd})-(\ref{aRd}). This of course is valid only, if we take a non-phantom matter $(m>0)$ into account.
Using (\ref{thetaav}), we then have
\bea
&&<\dot{\theta}>_{SFS,L} = \lim_{\genfrac{}{}{0pt}{}{t_0 \to -t_B}{t_1 \to 0}} -3n \frac{\int_{t_0}^{t_1} (-t)^{3n-2} dt}{\int_{t_0}^{t_1} (-t)^{3n}dt}
\label{sfsavtL}
\\
&=& \lim_{\genfrac{}{}{0pt}{}{t_0 \to -t_B}{t_1 \to 0}} - 3n \frac{3n+1}{3n-1} \frac{(-t_1)^{3n-1} - (-t_0)^{3n-1}}{(-t_1)^{3n+1} - (-t_0)^{3n+1}} \to \frac{1}{t^2_B}~~, \nonumber
\eea
\bea
&&<\dot{\theta}>_{SFS,R} = \lim_{\genfrac{}{}{0pt}{}{t_0 \to 0}{t_1 \to t_C}} 3n \frac{\int_{t_0}^{t_1} t^{3n-2} dt}{\int_{t_0}^{t_1} t^{3n}dt}
\label{sfsavtR}
\\&=& \lim_{\genfrac{}{}{0pt}{}{t_0 \to 0}{t_1 \to t_C}} 3n \frac{3n+1}{3n-1} \frac{t_1^{3n-1} - t_0^{3n-1}}{t_1^{3n+1} - t_0^{3n+1}} \to \frac{1}{t^2_C}~~, \nonumber
\eea
independently of $\delta$, $a_s$, $t_B$, and $t_C$ and for $1<n<2$. This last condition ($1<n<2$) guarantees that a sudden singularity appears, and so we deal with a type II model (finite scale factor and the energy density, divergent pressure). However, for a type III model with a finite scale factor singularity (both pressure and the energy density divergent), one has $0<n<1$ and the situation changes. Evidently, the averages (\ref{sfsavtL}) and (\ref{sfsavtR}) blow-up for $0<n<1/3$. In that sense a finite scale factor singularity is stronger than a big-bang singularity. Of course this is not the case for generalized sudden future singularity models for which $n>2$ \cite{SFS,APS2010}.

As for the pressure and the energy density averages, according to (\ref{peav}) and (\ref{roav}) we have that
\bea
<p>_L &=& \lim_{\genfrac{}{}{0pt}{}{t_0 \to -t_B}{t_1 \to 0}} - n (3n-2)  \frac{\int_{t_0}^{t_1} (-t)^{3n-2} dt}{\int_{t_0}^{t_1} (-t)^{3n}dt}~~,
\label{sfsavpL}\\
<p>_R &=& \lim_{\genfrac{}{}{0pt}{}{t_0 \to 0}{t_1 \to t_C}} - n (3n-2)  \frac{\int_{t_0}^{t_1} t^{3n-2} dt}{\int_{t_0}^{t_1} t^{3n}dt}~~,
\label{sfsavpR}\\
<\varrho>_L &=& \lim_{\genfrac{}{}{0pt}{}{t_0 \to -t_B}{t_1 \to 0}} 3n^2  \frac{\int_{t_0}^{t_1} (-t)^{3n-2} dt}{\int_{t_0}^{t_1} (-t)^{3n}dt}~~,
\label{sfsavrL}\\
<\varrho>_R &=&
\lim_{\genfrac{}{}{0pt}{}{t_0 \to 0}{t_1 \to t_C}} 3n^2  \frac{\int_{t_0}^{t_1} t^{3n-2} dt}{\int_{t_0}^{t_1} t^{3n}dt}~~.
\label{sfsavrR}
\eea
From (\ref{sfsavpL})-(\ref{sfsavrR}), after taking appropriate limits, we can easily conclude that the spacetime averages are also finite.

It is possible to show that a $w-$singularity \cite{wsing} has also a finite spacetime average. In order to do that, one can take the Taylor series of the scale factor $a(t)$ at $t_s$, as given in Ref. \cite{LFJ}
\be
a(t) = a_s + \sum_{i=3}^{\infty} a_i (t_s - t)^i = a_s + a_3 (t_s - t)^3 + \cdots~~,
\label{waseries}
\ee
where $t_s$ is a $w-$singularity time, and $a_s$, $a_i$ are constants. The derivatives of (\ref{waseries}) are
\bea
\dot{a} &=& - \sum_{i=3}^{\infty} i a_i (t_s - t)^{i-1} = - 3 a_3 (t_s - t)^2 + \cdots,
\label{wdaseries} \nonumber\\
\ddot{a} &=& \sum_{i=3}^{\infty} i(i-1) a_i (t_s - t)^{i-2} = 6 a_3 (t_s - t) + \cdots,
\nonumber \label{wddaseries}
\eea
so that the barotropic index which reads as
\be
\gamma + 1 = w = -\frac{1}{3} - \frac{2}{3} \frac{a\ddot{a}}{\dot{a}^2} \sim \frac{1}{t_s - t}~~,
\label{windex}
\ee
blows-up for $t=t_s$. Having (\ref{waseries})-(\ref{wddaseries}), one may calculate spacetime averages of the acceleration scalar, the pressure and the energy density as follows
\bea
<\dot{\theta}>_w &=& \frac{- 9a_s a_3 t_s^2 + \cdots - 9 a_sa_3 t_s^3 + \cdots}{
a_s^3 t_s + \cdots - \frac{1}{4} a_3^3 t_s^4 + \cdots}~~,
\label{wtheta}\\
<\dot{p}>_w &=& \frac{-6a_s a_3 t_s^2 + \cdots + 3a_s a_3 t_s^3 + \cdots}{
a_s^3 t_s + \cdots - \frac{1}{4} a_3^3 t_s^4 + \cdots}~~,
\label{wp}\\
<\dot{\varrho}>_w &=& \frac{-9a_sa_3^2t_s^3 + \cdots}{
a_s^3 t_s + \cdots - \frac{1}{4} a_3^3 t_s^4 + \cdots}~~,
\label{wvarrho}
\eea
where we have taken the limits from $t_0 = 0$ to $t_1 = t_s$. One easily sees from (\ref{wtheta})-(\ref{wvarrho}) that these averages are finite.

\section{Conclusion}
\setcounter{equation}{0}
\label{conclusion}

We have shown that the spacetime average of the standard matter (barotropic index $\gamma = w+1 > 0$) big-bang models is zero.
On the other hand, we have found that the phantom matter ($\gamma <0$) spacetime average is infinite. This may suggest that the
appearance of the phantom-driven big-rip singularities can be connected with a blow-up of the spacetime average while this is not the
case for standard big-bang singularities. In other words, bearing in mind these tools, phantom-driven singularities
are stronger singularities than big-bangs and big-crunches.

We have also shown that for sudden future singularities (and their generalizations) the spacetime average is zero while for finite
scale factor singularity (which allows both the energy density and the pressure to diverge) this average for some values of the model
parameters can be infinite. In that sense finite scale factor singularities may be considered  stronger singularities than big-bangs
and big-crunches. We have proven that $w-$singularities have finite spacetime average, too.

The final conclusion is that it is not obvious to find the proper measure/indicators for the appearance of singularities in the universe.

\section{Acknowledgements}

 I acknowledge the hospitality of Yukawa Institute in Kyoto, Japan during Gravity and Cosmology 2010 Workshop, where this paper was initiated. This work was also supported by the National Science Center grant No N N202 3269 40.

\end{document}